\documentclass[prl,twocolumn,superscriptaddress]{revtex4}%

\usepackage{amsfonts}
\usepackage{amsmath}
\usepackage{amssymb}
\usepackage{graphicx}%
\begin{document}
\preprint{cond-mat/0302431} %
\title{Evidence for Unconventional Superconductivity in the Non-Oxide Perovskite $\mathrm{MgCNi_3}$
from Penetration Depth Measurements}

\author{R.~Prozorov}
\affiliation{Department of Physics \& Astronomy and USC NanoCenter, University of South Carolina,
712 Main St, Columbia, SC 29208.}

\author{A.~Snezhko}
\affiliation{Department of Physics \& Astronomy and USC NanoCenter, University of South Carolina,
712 Main St, Columbia, SC 29208.}

\author{T.~He}
\affiliation{Department of Chemistry and Princeton Materials Institute, Princeton University, Princeton, NJ 08544.}

\author{R.~J.~Cava}
\affiliation{Department of Chemistry and Princeton Materials Institute, Princeton University, Princeton, NJ 08544.}



\begin{abstract}
The London penetration depth, $\lambda(T)$, was measured in polycrystalline powders of the non-oxide
perovskite superconductor $\mathrm{MgCNi_3}$ by using a sensitive tunnel-diode resonator technique. The
penetration depth exhibits distinctly non s-wave BCS low-temperature behavior, instead showing quadratic
temperature dependence, suggestive of a nodal order parameter.
\end{abstract}
\volumeyear{year} \volumenumber{number} \issuenumber{number}
\date{20 February 2003}

\maketitle

Identification of the symmetry of the order parameter in superconductors is one of the most challenging
experimental problems in distinguishing conventional from unconventional superconductivity. Theoretically,
there are several possibilities, including the Bardeen, Cooper and Schrieffer (BCS) s-wave \cite{bardeen} and
unconventional p- and d-wave pairing scenarios \cite{annett,scalapino,brandow,hirschfeld}. Each particular
symmetry imposes constraints on the possible mechanism of electron-electron pairing. Determination of the
pairing type, however, is often difficult. Electron-doped high-$T_c$ cuprates, for example, were thought to
exhibit s-wave BCS behavior until recently shown to be d-wave superconductors \cite{prozorov,tsuei}. The
recently discovered \cite{he} non-oxide perovskite superconductor $\mathrm{MgCNi_3}$ is especially important,
because it is viewed as a bridge between high-$T_c$ cuprates and conventional intermetallic superconductors.
This material is close to a magnetic instability on hole doping, and it is therefore natural to ask whether
an unconventional pairing mechanism might be operating \cite{rosner}. The absence of good single crystals and
oriented films does not allow the use of phase-sensitive techniques \cite{vanharlingen,tsueiRMP} to probe
pairing symmetry, and therefore other methods must be employed. Thermal, magnetization, resistivity, nuclear
spin-lattice relaxation, and tunnelling studies of $\mathrm{MgCNi_3}$ have been reported
\cite{lin,walte,kinoda,mao,karkin,granada}. The low-temperature London penetration depth measurements
reported here provide new insight into the nature of the superconductivity in $\mathrm{MgCNi_3}$.

The current experimental situation is highly controversial. On one hand, evidence for conventional s-wave
behavior is found in specific heat measurements \cite{lin,walte}, although the authors disagree on the
coupling coupling strength. The nuclear spin-lattice relaxation rate $1/^{13}CT_1$, seems to exhibit behavior
characteristic of an s-wave superconductor \cite{singer}. Some tunnelling data support conventional s-wave
pairing \cite{kinoda}. On the other hand, a zero-bias conductance peak (ZBCP) attributed to Andreev bound
states has been observed, and it was argued that the observed ZBCP could not be due to intergranular coupling
or other spurious effects \cite{mao}. Nonmagnetic disorder introduced by irradiation was found to
significantly suppress superconductivity \cite{karkin}. Such suppression is not expected in materials with a
fully developed gap, and is a strong indication of an order parameter with nodes. Theoretical calculations
support this conclusion \cite{granada}. Furthermore, recent theoretical developments predict the possibility
of a unique unconventional state \cite{voelker}, which might reconcile apparently contradictory experimental
observations.

Previous studies conclude that more experimental data is needed in order to draw a conclusion regarding the
pairing symmetry in $\mathrm{MgCNi_3}$. It is very difficult to experimentally identify the non-exponential
contribution of low-energy quasiparticles due to the presence of nodes in the superconducting gap on the
Fermi surface. In the case of thermal measurements, this electronic contribution is masked by a large phonon
contribution. For electromagnetic measurements, sensitivity is typically a problem. Precise measurements of
the London penetration depth are therefore very important.

In this letter we report measurements of the magnetic penetration depth, $\lambda ( T )$, down to 0.4 K in
polycrystalline powders of the the non-oxide perovskite superconductor $\mathrm{MgCNi_3}$ ($Tc \approx 7.2$
K). The sample employed for this measurement was exactly the one of composition $\mathrm{MgC_{0.98}Ni_3}$
characterized by neutron diffraction \cite{amos}. The synthesis is described in detail in that publication.

In order to avoid artifacts related to possible inter-grain coupling, three different samples were prepared:
powder mixed in paraffin, powder mixed and solidified in low-temperature Stycast 1266 epoxy, and a pellet
sintered at room temperature and 2.5 GPa for 8 hours. All samples showed similar low-temperature behavior,
indicating no additional contribution from inter-grain coupling. We note that our previous measurements of
$\mathrm{MgB_2}$ powder of similar grain size gave results fully consistent with s-wave symmetry and are in
complete agreement with measurements performed on single crystals \cite{manzano}. In addition, a sample cut
from a polycrystalline niobium foil was measured for comparison.

\begin{figure}[tb]
\includegraphics[width=8.5cm]{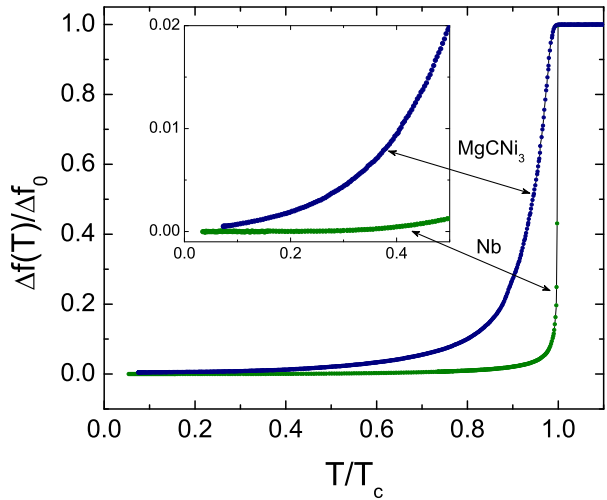}%
\caption{Magnetic penetration depth measured in zero external field in $\mathrm{MgCNi_3}$ (upper curves)
compared to polycrystalline Nb foil (lower curves). The inset shows the low-temperature behavior.}%
\label{fig1}%
\end{figure}

The penetration depth, $\lambda (T)$, was measured by using a $13\ MHz$ tunnel-diode driven LC resonator
\cite{resonator,prozorov3} mounted in a $^{3}He$ refrigerator. An external dc magnetic field ($0-6$ T) could
be applied parallel to the ac field ($\sim5$ mOe). The oscillator frequency shift $\Delta f=f(T)-f(T_{min})$
is proportional to the linear ac susceptibility and, therefore, to the change in the penetration depth,
$\Delta\lambda=\lambda(T)-\lambda(T_{min})$ \cite{prozorov3}. At low temperatures, $\Delta f=-\Delta f_0
\Delta \lambda/R$, where $\Delta f_0$ is the total frequency shift when the perfectly diamagnetic sample is
inserted into an empty resonator, and $R$ is the characteristic sample size \cite{resonator,prozorov}. In the
case of powders, the observed frequency shift is the sum of contributions from individual grains. To verify
this assumption, we solved numerically the two-dimensional London equation for different assemblies of grains
of various (including nonanalytic) shapes, grain-grain distances, and $\lambda/R$ ratios. The response was
always additive, with no noticeable interference effects. The computations were done with Femlab multiphysics
Toolbox \cite{femlab} in Matlab \cite{matlab}, as described in greater details elsewhere \cite{prozorov3}. A
similar experimental approach using sensitive bulk magnetization measurements on powder superconductors was
effectively employed to study penetration depth in high-$T_c$ cuprates \cite{panagopoulos}. It has also been
successfully used in tunnel-diode resonator measurements of $MgB_2$ powders \cite{manzano} and
polycrystalline wires \cite{prozorov2}.

Although we cannot extract the absolute value of the magnetic penetration depth (this would require knowing
grain shapes and sizes with the accuracy of $\lambda(0)$ itself \cite{prozorov4}), our technique provides a
very sensitive ($1$ part per $10^{10}$) detection of the \textit{change} in the penetration depth. By varying
temperature, $\Delta \lambda(T)$ is obtained. In all plots, $\Delta f(T)/\Delta f_0$, proportional to $\Delta
\lambda (T)$ through a calibration constant, is shown. The calibration constant depends on average grain size
and the number of grains per unit volume of a composite material, and is difficult to estimate reliably.
Importantly, it does not influence the temperature variation, which is the focus of this work.

\begin{figure}[tb]
\includegraphics[width=8.5cm]{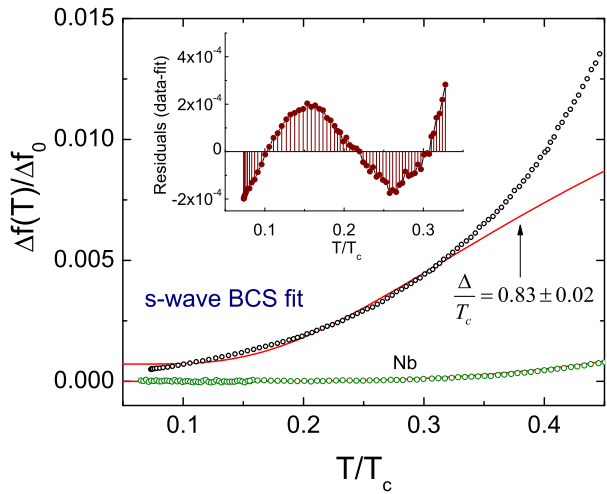}%
\caption{Upper curve: Best s-wave BCS fit for $\mathrm{MgCNi_3}$ using a standard expression (described in
the text) with $\Delta(0)/T_c$ as a free parameter. The lower curve shows Nb sample and a standard BCS fit.
Inset: residuals, $Data-Fit$, for the best BCS fit, which yields $\Delta(0)/T_c=0.83 \pm 0.02$.}
\label{fig2}%
\end{figure}

Clear evidence for a d-wave superconducting order parameter is linear temperature variation of the London
penetration depth, $\Delta \lambda(T)/\lambda(0) \approx \ln{2}/\Delta(0) T$ \cite{scalapino,hirschfeld}.
 In a conventional s-wave superconductor, on the other hand, an exponential
decay is expected for the penetration depth: $\Delta \lambda = \lambda (0) \sqrt{\pi \Delta
(0)/2T}\exp{(-\Delta (0)/T)}$ for $T \lesssim 0.32 T_c$ with $\Delta (0)/T_c=1.76$ \cite{annett,scalapino}.
Measurements on a non oriented powder mean that the result is averaged over all contributions
($\lambda_{a,b,c}$). Fortunately, $\mathrm{MgCNi_3}$ is isotropic and therefore we obtain values
characteristic for this material.

Figure~\ref{fig1} presents $\lambda(T)$ measured in $\mathrm{MgCNi_3}$ powder mixed in paraffin. The data is
compared with the measurements performed on a sample cut from a polycrystalline niobium foil. The niobium
data is fully consistent with the weak coupling s-wave BCS picture (in the entire temperature range). The
data for $\mathrm{MgCNi_3}$ also approach saturation on decreasing temperature. The magnetization measured on
a commercial magnetometer would show no temperature dependence in the low-temperature region. However, our
resolution is sufficient to study the low-temperature part. Apparently, the data obtained for
$\mathrm{MgCNi_3}$ is strikingly different from that of Nb.

Although the observed temperature dependence is obviously not exponential, it is instructive to attempt to
fit the data to the standard low-temperature BCS form with $\Delta(0)/T_c$ being a free parameter.
Figure~\ref{fig2} shows such a best fit, which clearly does not describe the data. In addition, the extracted
$\Delta(0)/T_c=0.83\pm0.02$ is too low. The inset shows the residual, $Data-Fit$, which reveals large
systematic deviation from the BCS behavior down to the lowest temperature.

\begin{figure}[ptb]
\includegraphics[width=8.5cm]{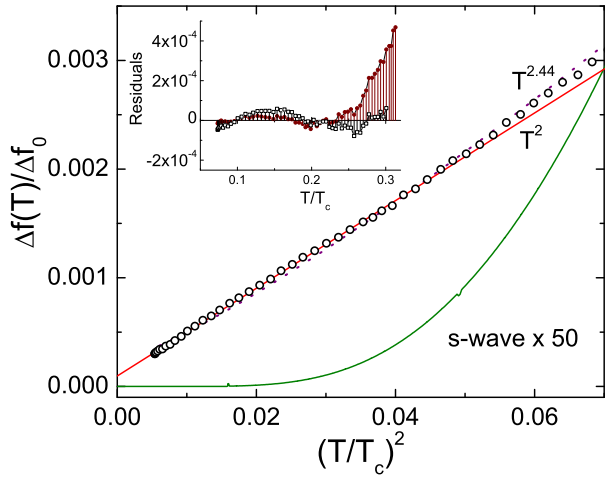}%
\caption{Penetration depth plotted versus $T^2$ compared to a standard s-wave curve. The fit to a pure $T^2$
behavior is shown by solid line. The fit to $T^n$ with $n$ as a free parameter is shown by dotted line.
Inset: residuals for fit with $n=2$ (solid symbols) compared to the residuals of $n=2.44$ fit. The vertical
scale of the inset is the same as in the inset of Fig.~\ref{fig2} for comparison.}%
\label{fig3}%
\end{figure}

The measured temperature dependence of $\Delta \lambda (T)$ is plotted versus $T^2$ in Fig.~\ref{fig3}. The
observed behavior is quite linear on this $T^2$ scale up to $T/T_c \approx 0.25$. The inset shows the
residuals plot, which confirms an overall good agreement of the fit with the experimental data. The residuals
plot scales in the insets to Fig.~\ref{fig2} and Fig.~\ref{fig3} have the same absolute ordinate scale for
easy visual comparison, showing the dramatically better power-law fit to the data. Also shown in
Fig.~\ref{fig3} is a fit to the power-law dependence, $\lambda(T) \sim T^n$ with the exponent $n$ as a free
parameter. The best fit gives $n \approx 2.44$, however, this is fit-range dependent. The obtained values of
$n$ decrease upon reduction of the fit-range and approach $n=2$ below $T/T_c \approx 0.25$, which is another
indication of the robustness of the inferred $\lambda(T) \sim T^n$ behavior. The residuals of the $n=2.44$
fit are compared to the $n=2$ residuals in the inset to Fig.~\ref{fig2}.

In a clean d-wave superconductors, a linear temperature dependence of $\Delta
\lambda(T)/\lambda(0)=\ln{2}T/\Delta(0)$ is predicted \cite{scalapino,hirschfeld} and observed \cite{hardy}.
However, this behavior is not expected in our case of microcrystalline powder with natural grain surface
roughness. In such a case, temperature dependence resulting from impurity scattering provides a more
plausible model, where a quadratic temperature variation of $\Delta \lambda(T)$ is expected
\cite{scalapino,hirschfeld,hardy}. There is an alternative explanation for $T^2$ behavior in a d-wave
superconductor. The divergence of the effective coherence length, $\xi=h v_F/\pi \Delta (\textbf{k})$ (where
$v_F$ is the Fermi velocity), near the nodes of a d-wave order parameter yields $\Delta \lambda (T) \sim\
T^2$ due to nonlocal electrodynamics \cite{kosztin}. Nonlocality is predicted to arise below $T_{nonlocal}
\approx \xi (0) \Delta (0)/\lambda (0)$, where $\xi(0) $ is the coherence length at zero temperature. In
$\mathrm{MgCNi_3}$ $T_{nonlocal} \approx$ 0.05 $T/T_c$, estimated using reported superconducting parameters
\cite{mao}. Since we observe quadratic temperature dependence up to roughly $T/T_c=0.25$, nonlocality is
unlikely to explain the observed behavior.

\begin{figure}[ptb]
\includegraphics[width=8.5cm]{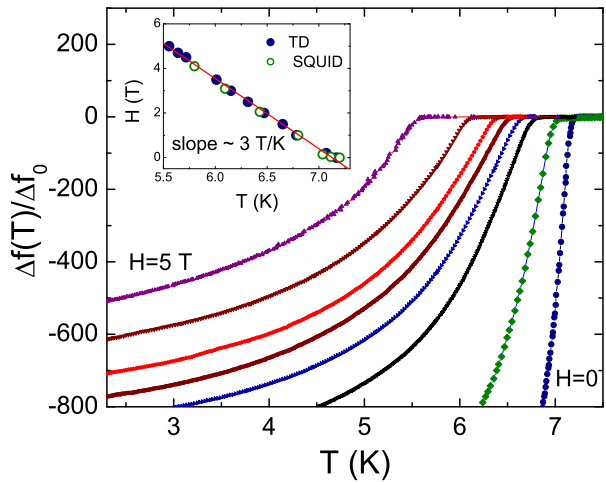}%
\caption{$\lambda(T)$ measured at different values of the external magnetic field, from $H=0$ to $H=5$ T. The
inset shows onset of superconductivity in tunnel diode measurements (closed symbols) compared to
SQUID measurements (open symbols).}%
\label{fig4}%
\end{figure}

Another possibility that might results in apparently non s-wave behavior of $\lambda(T)$ would be to have a
significant distribution of transition temperatures, $T_c$, due to inhomogeneities in chemical composition.
However, our numerical solution in the framework of the weak-coupling s-wave BCS theory indicates that in
order to mimic the $T^2$ behavior observed, the sample would have to contain a linear probability
distribution of $T_c$s extending from 7.2 to 0 K. This kind of distribution is chemically unfeasible, and, in
addition, is impossible for $\mathrm{MgC_xNi_3}$ because the perovskite phase becomes chemically unstable at
a minimum $T_c$ of 2.5 K \cite{amos}. The absence of phases with $T_c$s below 2.5 K means that what appears
to be non-BCS behavior cannot be induced in the low temperature range of interest here by chemical
inhomogeneity. Finally, there is no indication of chemical inhomogeneity induced broadening in the neutron
diffraction pattern \cite{amos} nor in the observed superconducting transition (see Fig.~\ref{fig4}),
indicating that the observed $T^2$ behavior cannot have a chemical origin.

The interpretation of our data in terms of a particular sureconductivity mechanism is further complicated by
the fact that some reports suggest that $\mathrm{MgCNi_3}$ is a multiband superconductor in which nontrivial
interband coupling may reconcile existing s-wave observations with unconventional superconductivity.
Calculations by Voelker and Sigrist \cite{voelker} performed along these lines call for new experimental
data, in particular penetration depth measurements. We hope results reported here will motivate further
theoretical study.

Figure \ref{fig4} shows measurements of the penetration depth at various values of the external DC magnetic
field. The overall behavior suggests weak pinning - the screening strength reduces due to a rapid increase of
the Campbell penetration depth. By measuring the onset of superconductivity at different fields, the
$H_{c2}(T)$ dependence can be reconstructed. The inset to Fig.~\ref{fig4} shows the onset temperature
compared to the onset temperature obtained by using \textit{Quantum Design} MPMS magnetometer. The good
agreement is independent evidence that our results, obtained on a 13 MHz resonator, are not introducing
undesirable frequency effects. From the measurements of the upper critical field, we obtain $dH_{c2}/dT
\approx 3$ T/K, which is consistent with previous measurements \cite{li,karkin}.

In conclusion, we have presented measurements and detailed experimental analysis of the London penetration
depth in the non-oxide perovskite superconductor $\mathrm{MgCNi_3}$. Our results show clear evidence for the
quadratic temperature variation of $\lambda(T)$ at temperatures below $\approx 0.25 T_c$. This behavior
indicates the presence of low-energy quasiparticles, and therefore unconventional non s-wave
superconductivity. It is consistent with d-wave pairing in the presence of strong impurity scattering, but
other nonconventional mechanisms may be implied.

Acknowledgements: Work at USC was supported by the NSF/EPSCoR under Grant No. EPS-0296165. Work at Princeton
was supported by the Department of Energy, grant No. DE-FG02-98-ER45706.

\end{document}